\documentclass[aip,apl,amsmath,preprint,endfloats]{revtex4-1}
\usepackage{graphicx}
\usepackage{dcolumn}

\begin{document}


\title{Influence of the substrate on the diffusion coefficient and the momentum relaxation in graphene: the role of surface polar phonons} 



\author{R. Rengel}
\email[Electronic mail: ]{raulr@usal.es}
\thanks{The following article appeared in Appl. Phys. Lett. 104, 233107 (2014) and may be found at http://scitation.aip.org/content/aip/journal/apl/104/23/10.1063/1.4882238}
\affiliation{Department of Applied Physics, University of Salamanca, Salamanca, 37008, Spain}


\author{E. Pascual}
\affiliation{Department of Applied Physics, University of Salamanca, Salamanca, 37008, Spain}

\author{M. J. Mart\'in}
\affiliation{Department of Applied Physics, University of Salamanca, Salamanca, 37008, Spain}

\date{\today}

\begin{abstract}
Knowing the influence of the substrate type on the diffusion coefficient and the momentum relaxation in graphene is of great importance for the development of new device models specifically adapted to the peculiarities of this material. In this work, the influence of surface polar phonons at low and high electric fields is evaluated by means of ensemble Monte Carlo simulations for several types of substrates. The results show that at low fields surface polar phonons have a major role on reducing the scattering time, breaking the correlation of velocity fluctuations and degrading the diffusion coefficient. At high fields the differences with regard to suspended samples in terms of diffusivity and momentum relaxation tend to reduce, providing at the same time larger saturation velocities, particularly for h-BN.
\end{abstract}


\maketitle 


Monolayer graphene presents outstanding electronic properties, including elevated intrinsic mobilities at room temperature, \cite{Bolotin08b} well beyond the mobility values of traditional semiconductors. However, the presence of a supporting dielectric substrate significantly reduces the low-field mobility in graphene due to the influence of surface polar phonon (SPP) scattering.\cite{Chen08, Fratini08, Konar10, Perebeinos10} While the low-field mobility is reduced, the saturation velocity can be enhanced in supported graphene samples due to the strongly anisotropic nature of SPP scattering.\cite{Li10} SPP scattering can be also determinant in graphene channels, strongly affecting the current saturation. \cite{Perebeinos10}

In this work we present a Monte Carlo study of the influence of SPP scattering on the diffusion coefficient and the momentum relaxation dynamics in graphene on several substrate types (HfO$_2$, SiO$_2$, SiC and h-BN). The knowledge of the diffusion coefficient in the case of graphene is of great importance for the development of drift-diffusion models for device simulations, \cite{Ancona10} and it also helps to provide an accurate discussion of the physical processes governing the performance of injection terahertz (THz) graphene lasers. \cite{Ryzhii13} Moreover, the investigation of carrier relaxation dynamics and the influence of SPPs is critical for the analysis of the time response of supported graphene,\cite{Price12,Low12} since substrate surface phonons provide pathways for energy relaxation close to the Fermi surface.\cite{Tielrooij13} It is important to remark that, in degenerate materials, the diffusion coefficient ($D$) can not be directly obtained from the study of the velocity fluctuations of the ensemble, but from the fluctuations corresponding to an excess carrier population (small enough to be considered as a minor perturbation of the system and coupled to the background population by a carrier-carrier exchange mechanism) evolving according to a linearized Boltzmann transport equation; \cite{Thobel97}details about the Monte Carlo model and the procedure employed can be found in previous works focused on the study of suspended monolayer graphene. \cite{Rengel13paper,Rengel13}

\begin{table}
\caption{\label{tab:table1}Phonon energies and low and high frequency dielectric constants considered for each type of substrate. The values have been taken from references \cite{Konar10} (HfO$_2$, SiO$_2$ and SiC) and \cite{Perebeinos10} (h-BN).}
\begin{ruledtabular}
\begin{tabular}{lcccc}
& HfO$_2$ & SiO$_2$ & SiC & h-BN \\
\hline
$\hbar\omega_{SO1}$ (meV) & 19.42 & 55.98 & 116 & 101.7 \\
$\hbar\omega_{SO2}$ (meV) & 52.87 & 146.51 & - & 195.7 \\
$\kappa_{ox}^{0}$ & 22 & 3.9 & 9.7 & 5.09\\
$\kappa_{ox}^{\infty}$ & 5.03 & 2.5 & 6.5 & 4.1\\
\end{tabular}
\end{ruledtabular}
\end{table}

The scattering processes included in the model are intrinsic optical phonons, intervalley and intravalley acoustic phonons and SPP scattering. The scattering rates for acoustic and optical phonons considered are derived from \textit{ab initio} calculations. \cite{Borysenko10} The anisotropic, inelastic SPP scattering\cite{Fratini08,Konar10,Li10} is implemented in the model taking into account the Fr\"{o}hlich nature of this kind of interaction, by properly accounting for the angle dependence of the scattering probability integrand. The values of the phonon energies and the low and high frequency dielectric constants for the different substrate materials considered are shown in Table \ref{tab:table1}. The separation between the graphene layer and the underlying dielectric is considered to be 0.4 nm,\cite{Fratini08,Konar10,Li10} and the Thomas-Fermi screening factor is also taken into account in the model.\cite{Hwang07} In order to better evaluate the impact of remote phonon interactions, impurities or defects that could appear related to the presence of the underlying substrate are neglected.

\begin{figure}
\includegraphics{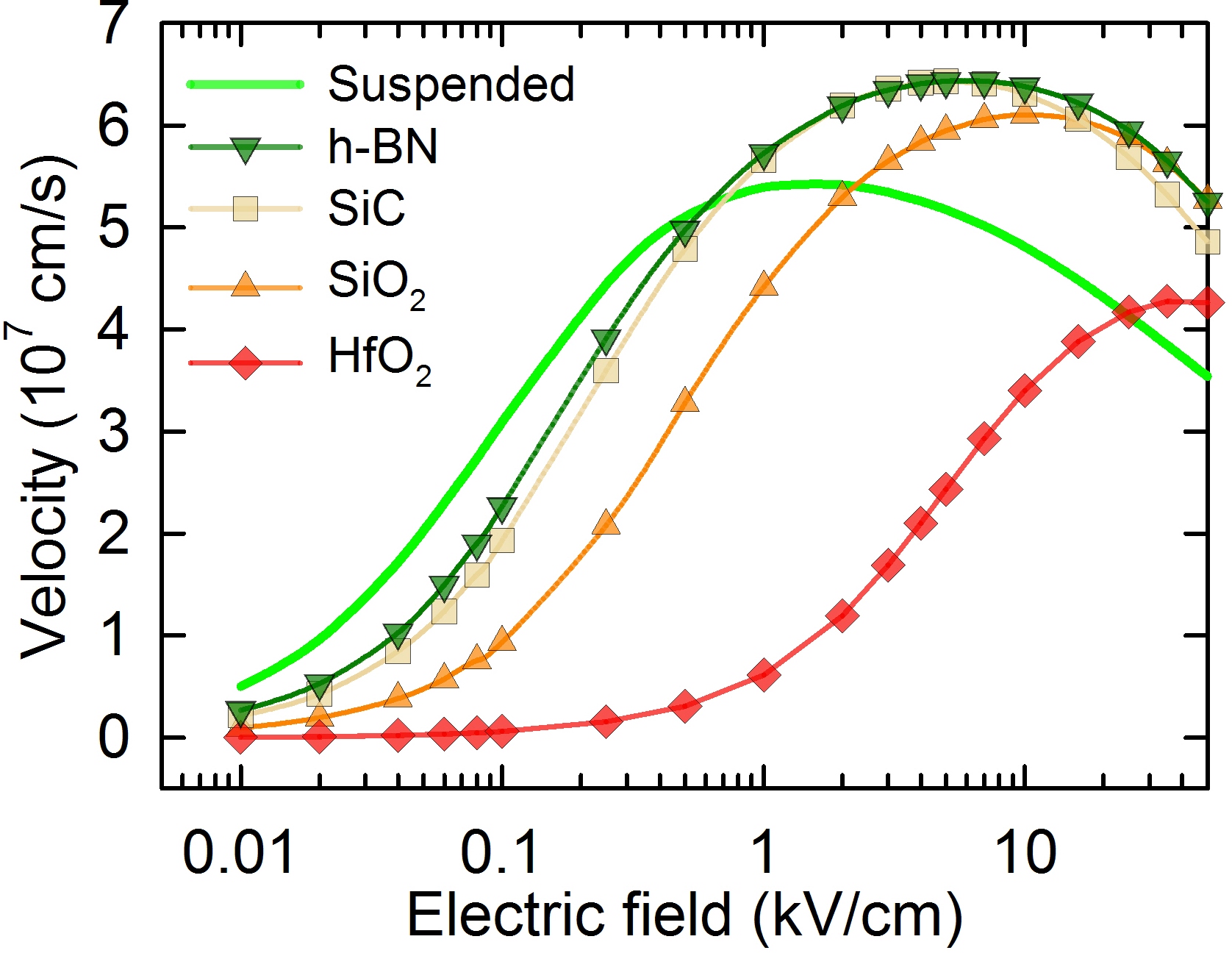}%
\caption{\label{Fig:vel-field}Drift velocity as a function of the applied electric field for suspended monolayer graphene and graphene on several different types of substrates. The carrier density is equal to $10^{12}$ cm$^{-2}$.}
\end{figure}

Figure \ref{Fig:vel-field} shows the drift velocity as a function of the applied electric field. The presence of SPP interactions significantly affects the velocity values and their dependence with the applied electric field, thus noticeably modifying the low-field mobility, the velocity saturation and the negative differential conductance at high electric fields. While the low-field mobility is significantly reduced, the consideration of SPP scattering is not translated into a reduced saturation velocity; on the contrary, higher velocities are consistently reached for all the supporting materials at high fields (even for HfO$_2$ over 25 kV/cm), which is in good agreement with studies based on displaced Fermi-Dirac distributions for solving the Boltzmann equation, \cite{Perebeinos10,DaSilva10} and with Monte Carlo simulations by other authors. \cite{Li10}

\begin{figure}
\includegraphics{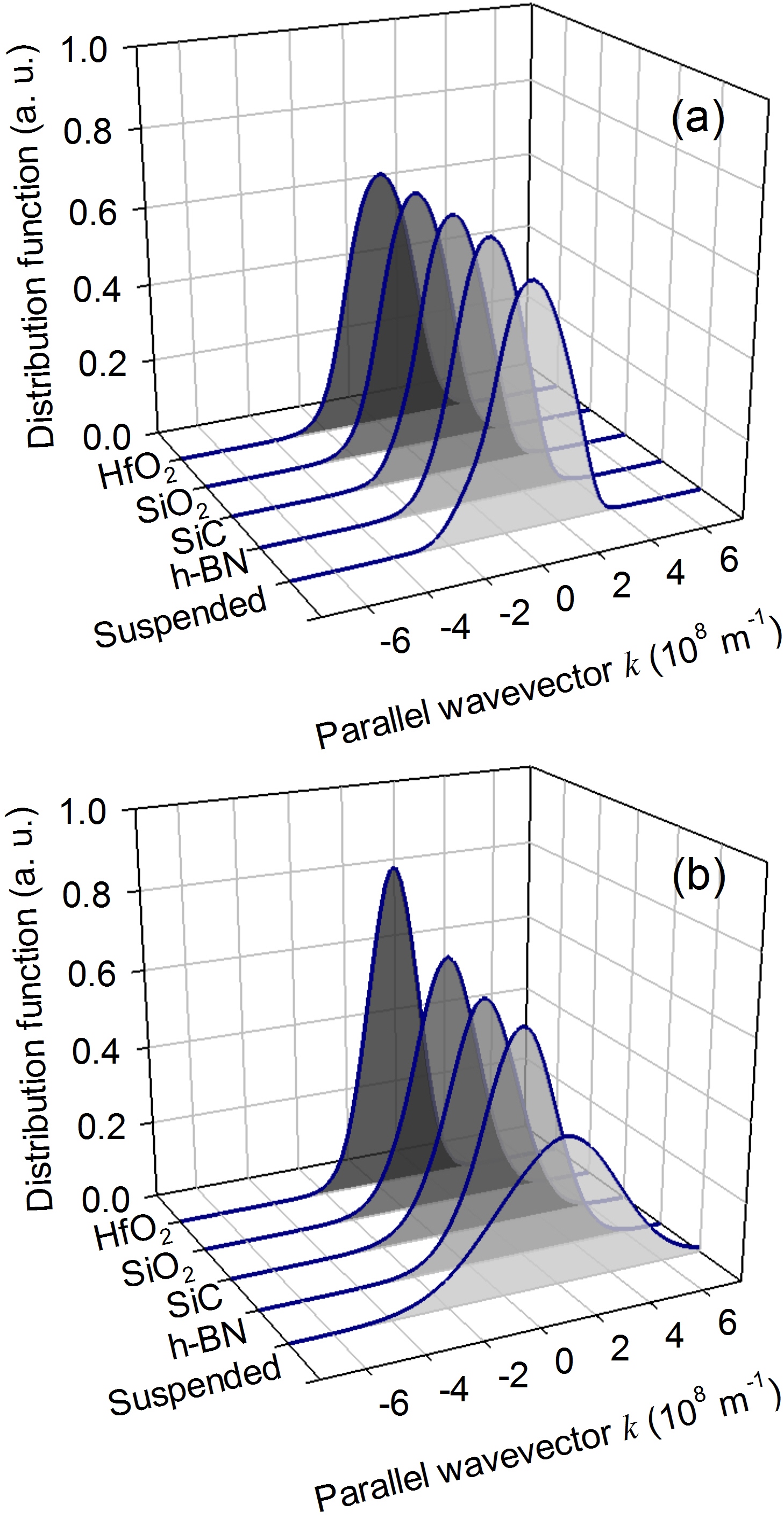}%
\caption{\label{Fig:fdist-k}Parallel momentum distribution at 0.1 kV/cm (a) and 10 kV/cm (b).}
\end{figure}

\begin{figure}
\includegraphics{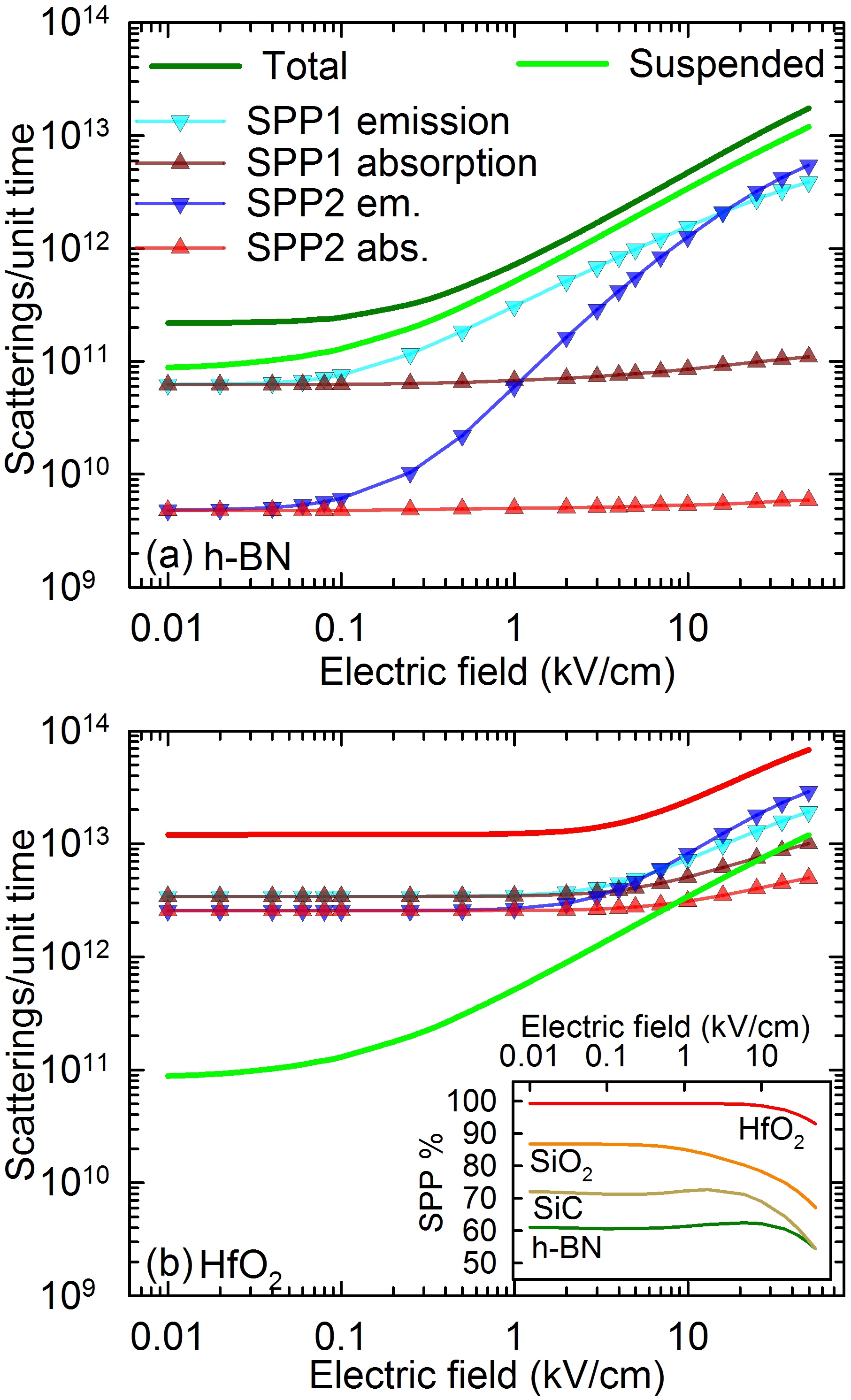}%
\caption{\label{Fig:scatterings}Number of scatterings per unit time in h-BN (a) and HfO$_2$ (b). The dark green in (a) and red in (b) lines account for the total number of scatterings, including intrinsic optical and acoustical phonons. The inset in (b) shows the total percentage of SPP scatterings. The results for the total number of scatterings in suspended graphene (light green lines) are shown for comparison purposes.}
\end{figure}

To analyze the origin of this behaviour, the parallel (in the direction of the applied electric field) momentum distribution is shown in Figure \ref{Fig:fdist-k} for the different cases under consideration, and the number of scatterings per unit time is presented for h-BN and HfO$_2$ in Figure \ref{Fig:scatterings}. At low fields (0.1 kV/cm, Figure \ref{Fig:fdist-k}(a)) the distribution shape is quite similar in all cases, with a progressively increasing small displacement towards higher $k$ values for HfO$_2$, SiO$_2$, SiC, h-BN and intrinsic graphene, in this order. The shift towards larger average $k$ is more relevant for suspended graphene, and implies also larger average energies. In the intrinsic material, at low fields long-wave (quasi-elastic) acoustic phonons are the dominant scattering mechanism,\cite{Rengel13paper} while for supported samples the largest number of scattering events corresponds to SPP interactions (see inset in Figure \ref{Fig:scatterings}(b)), with larger prevalence of absorption and emission processes corresponding to the lowest energy phonon for each substrate type (e.g., see Figure \ref{Fig:scatterings} for h-BN and HfO$_2$). The important activity of remote phonon interactions implies a much reduced total scattering time (larger number of scatterings per unit time) in supported samples at low fields. For example, while at 0.1 kV/cm the scattering time was found to be equal to 7.7 ps in intrinsic graphene, in graphene on HfO$_2$ it is equal to 0.08 ps. Although the scattering time does not correspond exactly to the momentum relaxation time due to the anisotropy of Fr\"{o}hlich interactions, it is a clear indicator of a much more rapid momentum relaxation dynamics in graphene on a supporting substrate caused by SPP interactions.

At high fields (Figure \ref{Fig:fdist-k}(b)), the momentum distribution presents noticeable differences in the cases under consideration. The distribution function shape becomes progressively narrower and displaced towards larger $k$ values for HfO$_2$, SiO$_2$, SiC, h-BN and intrinsic graphene, in this order, with also larger corresponding average carrier energies. In suspended graphene intrinsic optical phonons are the dominant scattering type at high electric field,\cite{Rengel13paper} while in supported samples SPP interactions are still clearly prevailing (inset of Figure \ref{Fig:scatterings}), in spite of a tendency to lose some influence at very large fields in benefit of intrinsic optical phonons. At high fields, the SPP interactions by secondary (more energetic) phonons become critical, and emission and absorption processes are no longer balanced, the emission being clearly more influent (Figure \ref{Fig:scatterings}). The inelasticity and anisotropy of SPP scattering yields to narrower momentum distribution functions with a reduced negative tail and less average carrier energy. This is the reason of the enhanced velocity values at high fields in supported samples as compared to suspended graphene.\cite{Li10} At high fields, the differences in the scattering time tend to reduce. For example, at 10 kV/cm the scattering time is equal to 0.29 ps in intrinsic graphene and 0.04 ps in graphene on HfO$_2$ (the material for which SPPs have the strongest influence).

The analysis of the correlation of instantaneous velocity fluctuations is critical for the investigation of diffusion processes.\cite{Jacoboni89} Although in supported graphene SPPs account for the most part of scatterings suffered by the electrons, individual SPP scattering events fail to completely reverse the momentum orientation: the occurrence of a SPP scattering interrupts the ballistic increase of the momentum, but in general it tends to not severely change its orientation. Consequently, velocities get closer to the physical limit, the Fermi velocity, for longer times (not shown in the graphs), since in graphene the velocity is directly linked to the momentum orientation and not to its module. However, the larger frequency of occurrence of scattering events breaks more rapidly the time correlation of instantaneous velocity fluctuations in the case of graphene on a supporting substrate: e.g., at 0.1 kV/cm the correlation time has been found to be approximately 10 ps in intrinsic graphene, 8 ps in h-BN and SiC, 5 ps in SiO$_2$ and 0.4 ps in HfO$_2$. Therefore, SPP plays a determinant role in the relaxation processes and in the reduction of the correlation time of instantaneous velocity fluctuations. At high fields, the correlation time is strongly reduced in all cases (below 1 ps), although slightly larger in suspended graphene.

\begin{figure}
\includegraphics{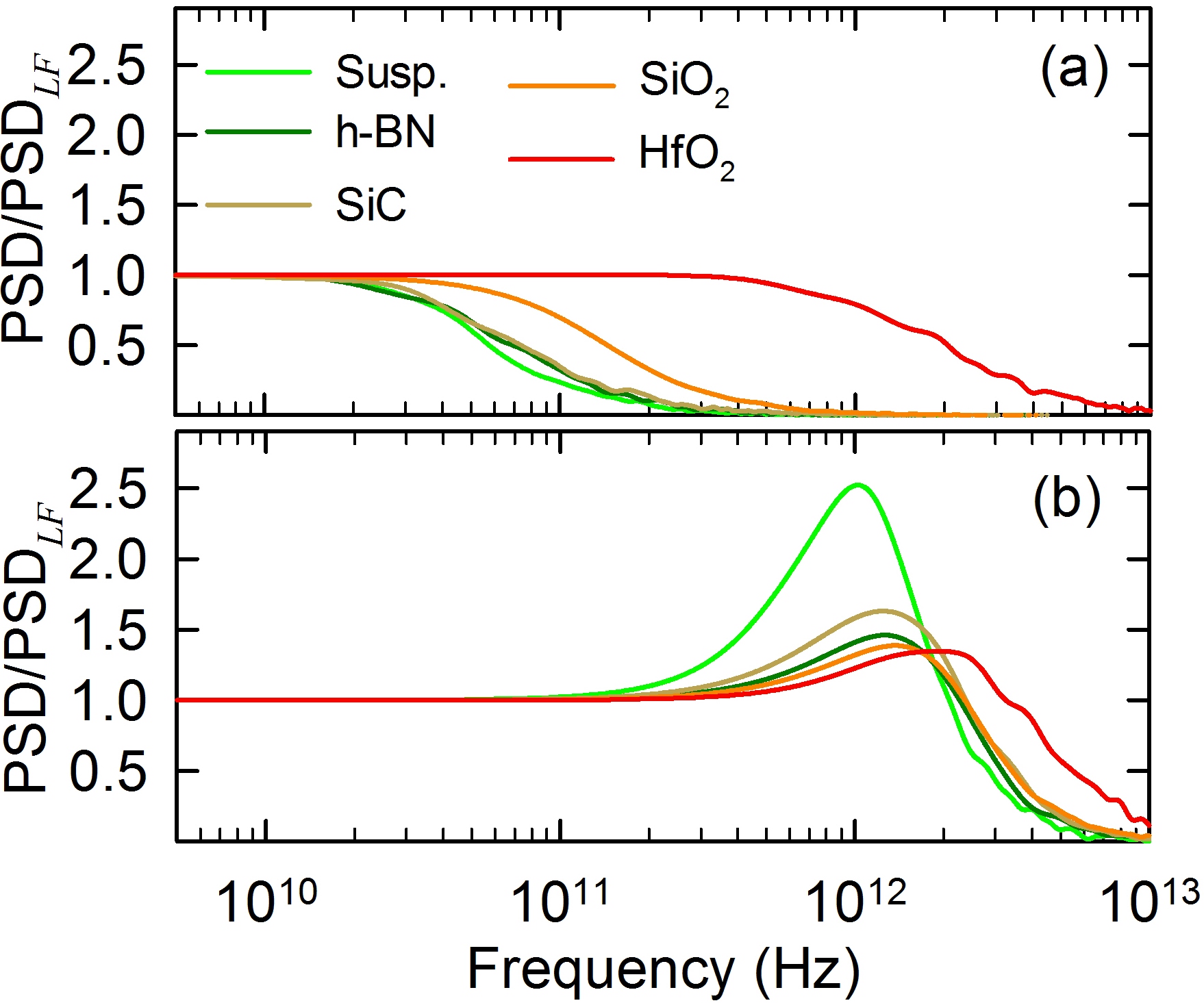}%
\caption{\label{Fig:dens-espec}Power spectral density (PSD) of instantaneous velocity fluctuations normalized by the low-frequency value, for an applied electric field equal to 0.1 kV/cm (a) and 10 kV/cm (b).}
\end{figure}

The power spectral density (PSD) (normalized to the low frequency value) of instantaneous velocity fluctuations is shown in Figure \ref{Fig:dens-espec} up to 10 THz. This is a relevant parameter for the analysis of diffusion processes, since it can be directly related to the diffusion coefficient by means of the Wiener-Kintchine theorem.\cite{Jacoboni89} In all cases, the PSD presents a white noise behaviour in the RF and microwaves frequency range. At low fields, after the white noise range a monotonic decay is observed: the corner frequency is larger for substrate samples, and particularly for those with the less energetic phonons (SiO$_2$ and HfO$_2$), in good agreement with the shorter correlation decay times and the preeminence of anisotropic scattering events in those cases. At high fields (Figure \ref{Fig:dens-espec}(b)), a maximum appears in the THz range, being more pronounced in the case of suspended graphene. The appearance of this peak is due to negative values of the velocity fluctuations correlation, which is strongly associated to the reversal of the momentum orientation induced by intrinsic optical phonons (change in the velocity sign); \cite{Rengel13paper} this effect is progressively less influent in h-BN, SiC, SiO$_2$ and particularly in HfO$_2$, where transport is more strongly influenced by anisotropic SPP scattering, which relaxes energy but fails to completely reverse the momentum.

\begin{figure}
\includegraphics{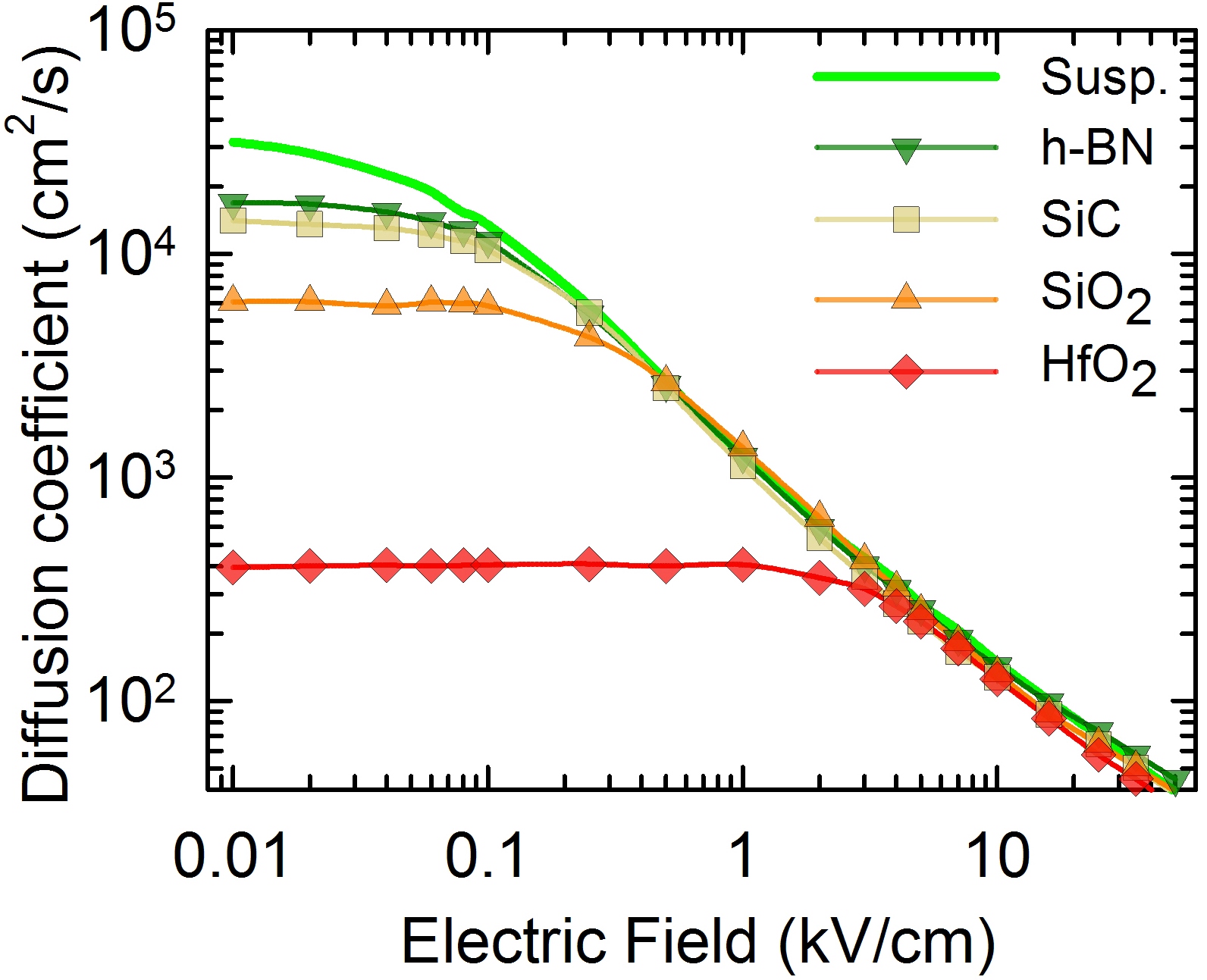}%
\caption{\label{Fig:diff-field}Parallel diffusion coefficient as a function of the applied electric field.}
\end{figure}

Figure \ref{Fig:diff-field} shows the parallel diffusion coefficient in the cases under study (the perpendicular diffusion coefficient, not shown in the graphs, shows a qualitatively similar behavior). The results have been obtained by means of the second central moment and the low frequency value of the PSD of velocity fluctuations, with an excellent agreement between both calculations.\cite{Rengel13paper} In the case of suspended graphene, values in excess of 31 000 cm$^2$/s are obtained at very low fields. The low-field values of the diffusion coefficient obtained for the different substrates are: 17 000 cm$^2$/s for h-BN, 14 000 cm$^2$/s for SiC, 6 100 cm$^2$/s for SiO$_2$ and 400 cm$^2$/s for HfO$_2$. The results obtained are consistent with experimental measurements (11 000 cm$^2$/s in epitaxial graphene on a Si-terminated 6H-SiC crystalline wafer surface and 5 500 cm$^2$/s in reduced graphene oxide samples on quartz substrates).\cite{Ruzicka10,Ruzicka12} As the electric field is augmented, the diffusion coefficient drops in all cases; however, it is noteworthy to mention that, the shorter the SPP energy is, the larger is the electric field range for which the diffusion coefficient remains practically constant; in the case of HfO$_2$ substrate, is keeps constant up to 1 kV/cm, while for h-BN it starts dropping at 0.04 kV/cm. For suspended graphene the constant $D$ region is practically absent. At high electric fields the differences between the substrates considered tend to decrease. In this case the intrinsic polar phonon activity in suspended samples becomes progressively more determinant (due also to a larger average energy at the same electric field) and effectively relaxes the momentum and velocity in a time scale closer to the case of SPP dominated transport, thus providing values of $D$ in suspended graphene similar to those obtained in supported samples.

This work has been supported by research project TEC2013-42622-R from the Ministerio de Economia y Competitividad.


%
%

%



\printtables
\printfigures
%

\end{document}